\documentclass[preprint2]{aastex6}

\usepackage{graphicx,natbib}
\usepackage{amsmath,amssymb}
\usepackage{color}
\usepackage{xspace}

\shorttitle{Flares on A-type stars}
\shortauthors{\v{S}vanda \& Karlick\'y}
\newcommand{\kepler}{{\it Kepler}\xspace}

\begin{document}

\title{Flares on A-type stars: Evidence for heating of solar corona by nanoflares?}
\author{Michal \v{S}vanda}
\affil{Astronomical Institute, Charles University in Prague, Faculty of Mathematics and Physics, V Hole\v{s}ovi\v{c}k\'ach 2, CZ-18000 Prague 8, Czech Republic}
\affil{Astronomical Institute (v. v. i.), Czech Academy of Sciences, Fri\v{c}ova 298, CZ-25165 Ond\v{r}ejov, Czech Republic}
\email{michal@astronomie.cz}
\author{Marian Karlick\'y}
\affil{Astronomical Institute (v. v. i.), Czech Academy of Sciences, Fri\v{c}ova 298, CZ-25165 Ond\v{r}ejov, Czech Republic}

\begin{abstract}
We analyzed the occurrence rates of flares on stars of spectral types K, G, F, and A,
observed by \kepler. We found that the histogram of occurrence frequencies of
stellar flares is systematically shifted towards a high-energy tail for A-type
stars compared to stars of cooler spectral types. We extrapolated the fitted
power laws towards flares with smaller energies (nanoflares) and made estimates
for total energy flux to stellar atmospheres by flares. We found that for
A-type stars the total energy flux density was at least 4-times smaller
than for G-stars. We speculate that this deficit in energy supply may explain
the lack of hot coronae on A-type stars. Our results indicate an
importance of nanoflares for heating and formation of the solar corona.
\end{abstract}

\section{Stellar flares across Hertzsprung-Russell diagram}

The Sun is considered a prototype star, where the availability of high-cadence
high-resolution observations together with long-term synoptic recordings allows
us to study all necessary details of stellar physics. The Sun is also known for
being a magnetically active star, with a variety activity phenomena connected
to this activity: sunspots, faculae, prominences, and flares. It is also highly
likely that the higher atmospheric layers with temperature inversion (the
chromosphere and the corona) do exist thanks to the magnetic fields
\citep{1996SSRv...75..453N}. Signs of magnetic activity on different stars are
also known from the literature \citep[see e.g. a recent paper
by][]{2016Natur.529..364S}. These signs consist of tentative evidence for
existence of (star)spots \citep[see][for a review]{2009A&ARv..17..251S},
prominence-like features \citep[e.g.][]{1996IAUS..176..449C} and flares
\citep[e.g.][]{1989SoPh..121..299P}, not to mention an evidence for activity
cycles \citep[e.g.][]{1978ApJ...226..379W,1995ApJ...438..269B}.

We will further focus on flares. Flares on the Sun seem to be a consequence of
the reconnection of the entangled magnetic field above the active region
\citep{2011LRSP....8....6S}. Flares on the Sun depict various energies with
smaller ones occurring frequently whereas the large flares are rare. Indications
for flares were also observed on other stars, first on young or
magnetically interacting stars, such as the RS CVn type, and magnetically
active M dwarfs. Flares on Sun-like stars were not seen until the availability
of the high-cadence high-precision photometry
\citep{2000ApJ...529.1026S}. \cite{2012Natur.485..478M} reported on
observations of large-energy flares on Sun-like stars recorded in the \kepler
light curves. This work was later extended by \cite{2013ApJS..209....5S} and
others.

Using a similar methodology, \cite{2012MNRAS.423.3420B} inspected over 10\,000
stars' light curves from \kepler public archives and
discovered Sun-like flares being common not only for Sun-like stars, but also
for stars of spectral types A to M. These flares had a intensity $0.001L_\star$ to $0.1L_\star$,
where $L_\star$ is a luminosity of the star, and a typical duration from
minutes to several hours. The typical energy was $10^{28}$~J for the M-F stars
and $10^{29}$~J for A stars. The authors also concluded that nearly all
stars in question vary at a low level with a period which is likely a
rotational period of the star. The flares on hotter-type stars likely do not
come from the cold companion, because the energy of these flares is typically
100-times larger than energy of flares expected on a cool companion. This
pioneering work was further elaborated later \citep{2015MNRAS.447.2714B}, where
it was found that incidence of flares on stars drops by only a factor of 4 from
K--M dwarfs to A--F stars. This was probably largely a selection effect:
contrast factor makes is easier to detect flares on cool stars, flares on hot
stars need to be more energetic in order to be detected. Flare energy was
strongly correlated with stellar luminosity and also correlated with stellar
radius roughly as a cube function, which can be understood in terms the length
of the loop undergoing reconnection being comparable to the radius of the star.
The larger the star is, the larger is the active region, hence larger volume of
the magnetic field (goes as a cube of the radius) and hence more energy
dissipated in the flare.

The origin of the flares on A-type stars is not well understood. A standard
model for Sun-like activity was developed for stars with near-surface
convection zone, where the interplay between the convection, rotation, and
meridional circulation redistributes and locally strengthens
the magnetic field frozen in stellar plasma. One could easily apply this
mechanism to stars cooler than roughly F5, where the near-surface convection
zone is believed to exist and supply enough mechanical energy, which is
converted to magnetism by the dynamo process. On a modelling side, first
self-consistent dynamo models of other than Sun-like stars were published
only recently. \cite{2013ApJ...777..153A} presented a 3-D magnetohydrodynamic model
for a $1.2~M_\odot$ rotating F-type star, where a dynamo effect appeared in
both convection and radiative zones. Magnetic field was organised into
large-scale structures. It formed strong toroidal bands (wreaths) of field with
intricate fibril structure and with enshrouding poloidal fields that served to
link the wreaths. The portion of the radiative zone played an active role in
both storing and building global-scale magnetic fields.

Stars hotter than F5 do not have an efficient mixing in near-surface layers,
they only host a thin convective shell \citep{2010ApJ...711L..35K} and a large
convection zone in their cores. This finding is consistent with findings of
\cite{2014ApJ...792...67C}, where they investigated the effects of the rotation
and sunspots coverage and effective temperature on occurrence rate of
superflares on G-M stars. The found that with increasing temperature the
incidence rate of superflare decreases, as dynamo action gets weaker due to the
decreased thickness of the sub-surface convection zone. The activity however
increases with an increasing rotation rate (inverse Rossby number), flare
energies are strongly correlated with rotation rate. So it would seem that
without the sub-surface convection zone no near-surface dynamo process should
exist. Yet, it seems that there is evidence for existence Sun-like magnetic
field even on these hotter stars.

The existence of magnetic fields on A-type stars was known for some time,
5--10\% of stars in the given mass range belong to the group of chemically
peculiar Ap stars with strong magnetic fields in a predominantly dipole
configuration with even kilogauss strengths. These fields are
considered to be fossil, as the decay time of these fields is estimated to be
longer than the lifetime of the star in question \citep{2004Natur.431..819B}.
\cite{2009A&A...500L..41L} reported on discovery of the magnetic field on Vega,
an A star which is not an Ap star. The characteristic level of the discovered
magnetic field was around 1~G disc integrated. Further investigation
\citep{2010A&A...523A..41P} reveals that this magnetic field is complex with a
large spot around the pole having a radial orientation of the magnetic field
(with a peak intensity around 7~G). This polar spot is accompanied by a small
number of magnetic patches in lower latitudes. Some of these small patches seem
to contain also an azimuthal geometry of the magnetic field. Over two years of
observations the spots seem to be quite stable. Today, there seems to be two
distinct groups of A stars when the magnetic field comes in question: Ap stars
with field intensity larger than 300~G and Vega-like stars with weak $\sim
1$-Gauss fields \citep{2007A&A...475.1053A,2014IAUS..302..338L}.

\begin{figure}
\includegraphics[width=0.47\textwidth]{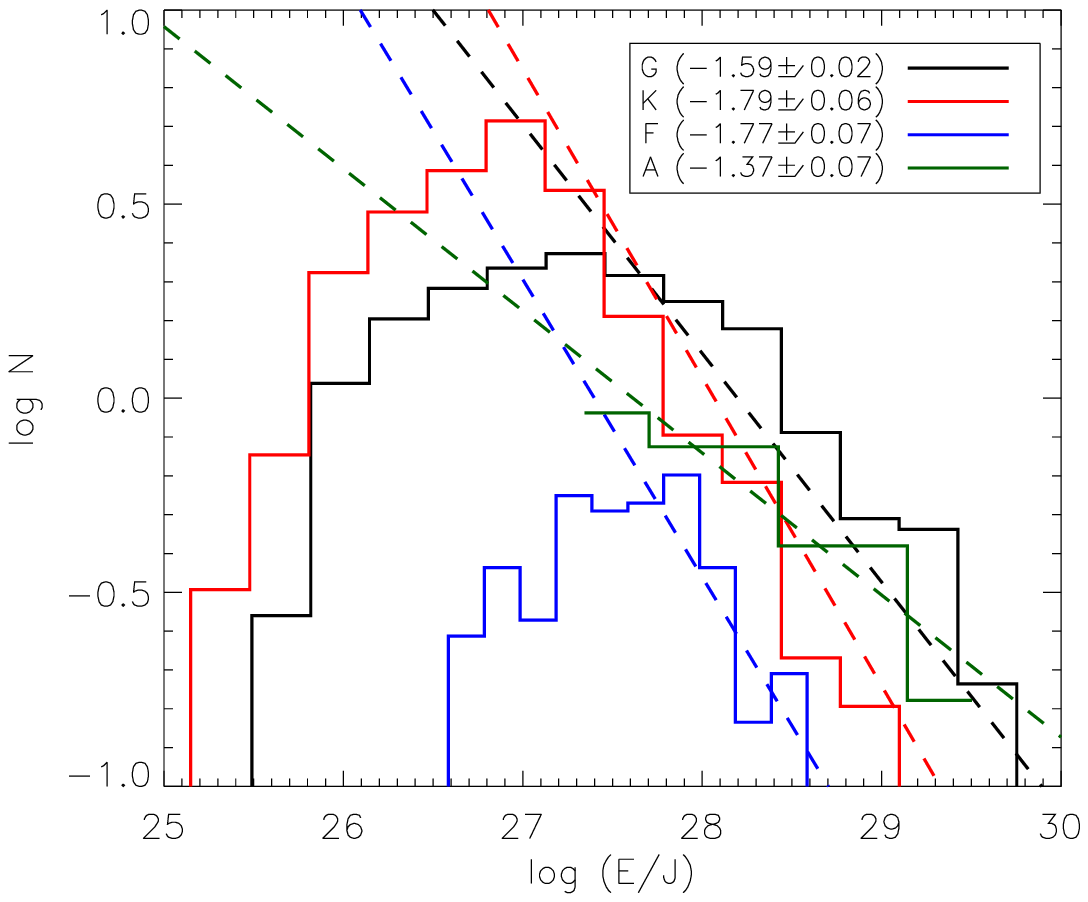}\\
\includegraphics[width=0.47\textwidth]{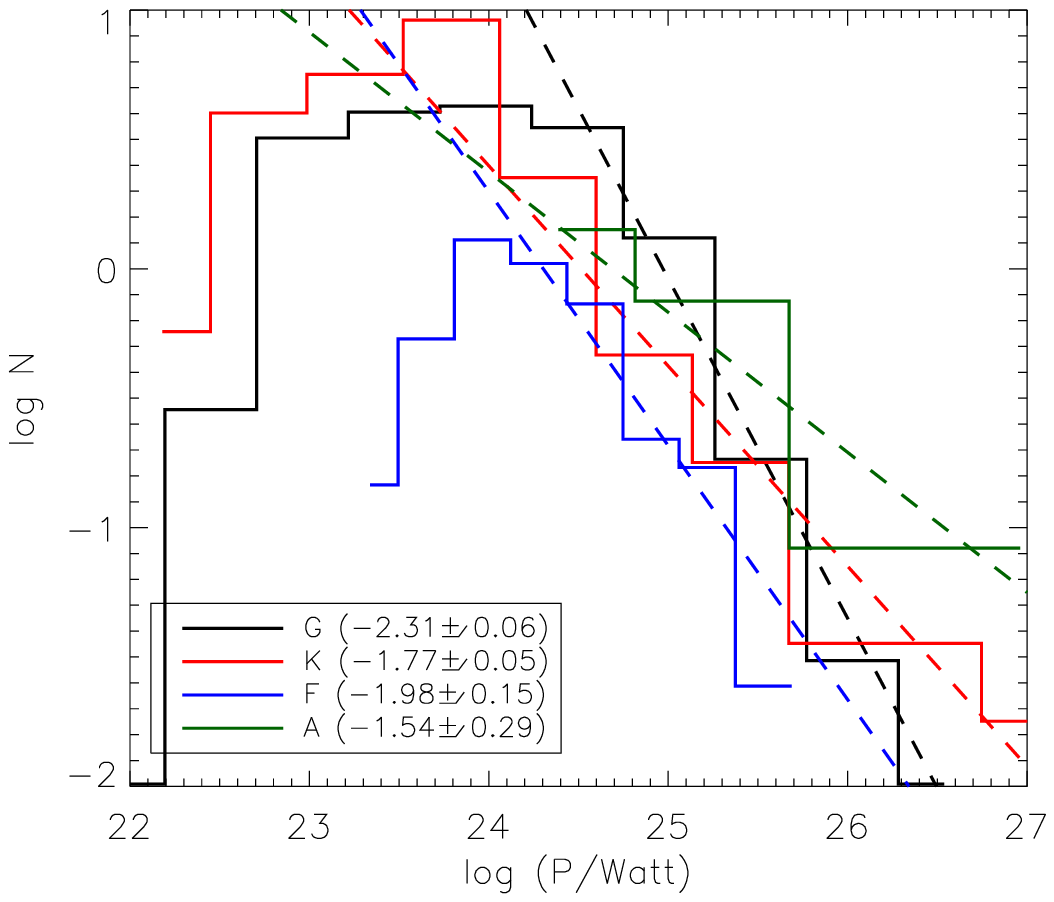}\\
\includegraphics[width=0.47\textwidth]{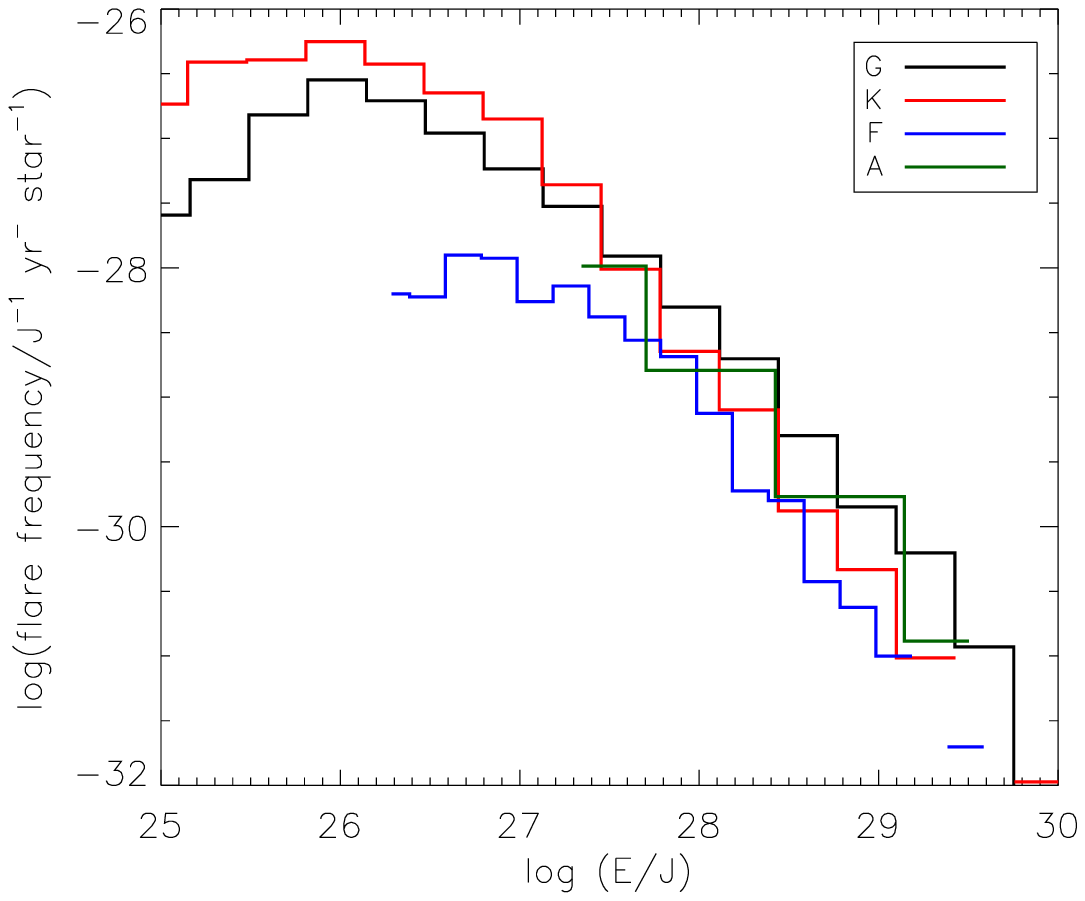}
\caption{\emph{Upper} -- Histograms of flare energies for various spectral types. Estimated power-law fits according to relation (\ref{eq:powerlaw}) are over-plotted. \emph{Middle} -- Analogous histograms for mean flare powers. \emph{Bottom} -- Histograms of occurrence rates of flares, which is comparable to similar plots seen in literature \citep[e.g.][]{2016arXiv160101132M}.}
\label{fig:histograms}
\end{figure}

\section{Statistics of stellar flares from Kepler SC data}
From \kepler short-cadence (SC) data \cite{2012MNRAS.423.3420B} detected
altogether 3140 flare events on 209 individual stars. These
flares have all the properties of solar flares, such as a rapid increase in
luminosity and a much slower decrease. We aimed to study the statistical
properties of the detected flares to assess the possible differences in the
physics related to their origin. Therefore we divided the set of 209 stars into
four corresponding spectral classes by their effective temperature, taken from
the Kepler Input Catalogue. The sample is dominated by G-type stars (98 stars,
1650 flares), whereas only 12 A-type stars with 28 recorded flares are present
in the sample. \cite{2012MNRAS.423.3420B} measured flare energies $E$ and their
durations $\tau$. By using a simple statistics one finds that on average, the
durations of the flares do not depend on the spectral type, a typical flare
duration is around 30 minutes. On the other hand, flares on A-type stars are by
one order more energetic ($E_{\rm mean} \approx 2\times 10^{29}$~J) than on F-K
stars ($E_{\rm mean} \approx 3\times10^{28}$~J) and the mean power $P=E/\tau$
is also larger ($P_{\rm mean}=6\times10^{25}$~W on A stars and $P_{\rm
mean}=4\times10^{24}$~W on F-K stars). The distribution of flare energies and
mean powers is highly asymmetrical, hence mean values are not representative,
however for the median value similar comparative relations are found.

Studies in the past
\citep[e.g.][]{1993SoPh..143..275C,1995PASJ...47..251S,2000ApJ...535.1047A}
using various data sets have shown that on the Sun, there a variety of flares
occurs differing by their total energy. Since the reconnection is essentially a
process with a self-organised criticality \citep{2016SSRv..198...47A}, the
occurrence rate of solar flares in energy might be described by a power law in a
form of
\begin{equation}
\frac{{\rm d} N(E)}{{\rm d} E} \propto E^\alpha,
\label{eq:powerlaw}
\end{equation}
where $N(E)$ is the number of flares at a given energy $E$ and $\alpha$ is the
index of the power law. For solar flares, $\alpha \sim -1.7$ based on various
studies. It has been shown \citep[e.g.][]{2016arXiv160101132M} that flares on
Sun-like G-type stars with large-energy flares (termed ``superflares'' in
literature) also follow a similar power-law dependence with an index close to
$-2$.

We performed similar analysis for flares detected also on A-, F-, and K-type
stars separately. The histogram of flare energies is displayed in
Fig.~\ref{fig:histograms} (the upper panel). The histograms for spectral types
in question were normalised by number of stars in the sample and the length of
\kepler SC runs. We fit the power laws to energies using the maximum
likehood method \citep{2007EPJB...58..167B} to obtain estimates for power-law
coefficients, including estimates of errors. This method is considered to
provide the best estimates of the power-law parameters compared to the fitting
of the histogram directly, which highly depends on an arbitrary selection of
the bins. The fitting was performed for the energies larger than the energy of
the maximum indicated on the corresponding histogram. There is a decreasing
number of flares with smaller energies, which is a selection bias due to the
detection limits \citep{2015MNRAS.447.2714B}.

The coefficients of the fitted power laws are similar for F-, and K-type stars,
larger for G-type stars, and significantly larger for A-type stars. We have to
point out that we cannot exclude that the less steep slope for the A stars is
due to the insufficient data sample at smaller energies. The histogram for A
stars does not contain decreasing branch towards lower energies, flares of energies lower than say $10^{27}$~J
are below the detection limit from the light curves and we do not expect this
issue to be resolved in a near future. Therefore it is not clear whether the
coverage of the lowest bin in the histogram for A stars is complete. In case it
is undersampled, the realistic value of the slope is steeper than estimated.
The planned photometric missions such as
TESS \citep{2015JATIS...1a4003R} or PLATO 2.0 \citep{2014ExA....38..249R} will not have photometric precision better than
one order, which will not allow to complete the low-energy part of flares in A stars. Both mission will potencially
help with an increased statistics by observing a larger sample of stars.
In the following, we assume that this effect is real and discuss possible implications.

We want to compare especially statistics of the flares on A-type stars to G-type
stars, where the respective slopes are $(-1.37\pm0.07)$ and $(-1.59\pm0.02)$.
These slopes that were fitted to values located on a decaying high-energy
branch of the histogram (the low-energy cut-offs were $2.2\times10^{27}$~J and
$4.2\times10^{27}$~J for A and G stars, respectively) are different to within $2.5\sigma$.

Similar comparison may be performed for the histograms of the mean powers (see Fig.~\ref{fig:histograms} -- middle panel), where again one sees a significantly larger power-law coefficient for A-type stars than for G-type stars. For comparison with previous works in the literature \citep[e.g.][]{2016arXiv160101132M} we plot also a histogram of occurrence rates by energy (Fig.~\ref{fig:histograms} -- the bottom panel).

\begin{table*}
\caption{Effect of the chosen threshold on descriptive statistics of the flares. For various threshold values the probabilities from Kolmogorov-Smirnof tests (always in pairs for both the histograms of energy and histograms of mean powers) are given. By a star we indicate situation, where with lesser than 5\% probability we make an error by assuming that the distribution function of both tested samples are different for both the energies and mean powers. }
\begin{tabular}{r|cccc}
Threshold [J]  & K-S A-G & K-S F-G & K-S K-G \\
\hline
$2\times 10^{27}$ & $\star$$0.02$, $3\times 10^{-5}$ & $\star$$7\times 10^{-4}$, $1\times 10^{-3}$ & $\star$$2\times 10^{-19}$, $7\times 10^{-40}$\\
$4\times 10^{27}$ & $\star$$0.12$, $4\times 10^{-3}$ & $\star$$5\times 10^{-4}$, $7\times 10^{-3}$ & $\star$$8\times 10^{-7}$, $4\times 10^{-9}$\\
$6\times 10^{27}$ & $\star$$0.19$, $6\times 10^{-3}$ & $2\times 10^{-5}$, $0.07$ & $\star$$0.02$, $0.05$\\
$8\times 10^{27}$ & $\star$$0.23$, $4\times 10^{-3}$ & $3\times 10^{-3}$, $0.11$ & $0.08$, $0.29$\\
$1\times 10^{28}$ & $\star$$0.03$, $5\times 10^{-3}$ & $0.04$, $0.83$ & $0.16$, $0.12$\\
$1.2\times 10^{28}$ & $\star$$0.04$, $1\times 10^{-3}$ & $0.11$, $0.78$ & $0.15$, $0.03$\\
$1.4\times 10^{28}$ & $\star$$0.05$, $2\times 10^{-3}$ & $0.42$, $0.57$ & $0.13$, $0.05$\\
\end{tabular}
\label{tab:K-S_tests}
\end{table*}
We performed Kolmogorov-Smirnov (K-S) tests to investigate whether the observed flare energy and mean power distribution functions differ for studied spectral types. The tests were applied to those parts of histograms, where the flare energies were larger than a chosen threshold to avoid observational biases in the low-energy part. Empirical distribution functions of A-, F-, and K-type stars were compared to a reference obtained from G-type stars. The choice of the threshold is purely arbitrary and affects the results. Therefore we performed the K-S testing for a set of threshold values. The results are summarised in Table~\ref{tab:K-S_tests}. It can be seen that when the distribution functions for energy and mean power of the flares are judged together, the histograms for A-type and G-type stars differ statistically significantly for all values of the threshold, whereas for the other spectral types when compared to G-type stars they might be considered the same especially in the region of large energies and mean powers.

\begin{table*}
\caption{Fundamental parameters of typical representatives of stars for each spectral type considered. In the first part are those derived from our sample of stars, in the second part the parameters are derived from the model}
\begin{tabular}{l|ccccc|cc}
\hline
Sp. type & $T_{\rm eff}$ [K] & $M$ [$M_\odot$] & $R$ [$R_\odot$] & $Z$ &  $P_{\rm rot}$ [days] & $R_{\rm MESA}$ [$R_\odot$] & $v_{\rm rot}$ [km\,s$^{-1}$]\\
\hline
 A   &   8570  &     2.29 &   3.28 & 0.017  &    5.54 & 3.42 & 31.2\\
 F   &   6360  &     1.41 &   2.38 & 0.014  &    2.59 & 2.33 & 46.8\\
 G   &   5490  &    1.00 &   1.51 & 0.014  &    4.05  &1.74 &22.0\\
 K   &   4200  &   0.61 &   0.76 & 0.025  &    5.68  & 0.66 & 5.9\\
\hline
\end{tabular}
\label{tab:stars}
\end{table*}

\section{Are A-type stars different from other types?}
To understand as to why the statistical distribution of flare energies seem to be
different for A-type stars from cooler types we construct a typical stellar
representative of each spectral class. We average fundamental parameters of the
stars of each spectral type to form a typical star in the sample. Since the
flares are in question, we would like our representatives to be rather
representative of the flaring stars, hence we weight the fundamental parameters
of each star contributing to the average by the number of recorded flares. This
way stars flaring often are weighted more than stars with only one or few
flares. The derived fundamental parameters of the representative stars is given
in Table~\ref{tab:stars}.

We construct stellar models for these representatives using the MESA
\citep{2011ApJS..192....3P} code. By trial-and-error approach we match the
following parameters of the model to those of the given representative:
effective temperature $T_{\rm eff}$, mass $M$, metalicity $Z$, and surface
rotation period $P_{\rm rot}$. For the input of the MESA code we use the model
for the Sun distributed with a MESA package as a template, on the upper
boundary we include a model of a grey atmosphere. In case there are multiple
solutions in $T_{\rm eff}$ (e.g. during the contraction phase and an expansion
on the main sequence), we use also the average radius and $\log g$ obtained
from the \kepler catalogue to distinguish between these two cases.

We used models of representative stars to assess the differences in their
atmospheric conditions that give us hints to explain the extended-tail in the
distribution of flare energies on A-type stars compared to the other types.
There are a few fundamental properties that affect the flare
magnetic reconnection. In case of solar flares the magnetic reconnection sets
on in the current sheets, where the electric resistivity increases suddenly
from classical (collisional) to anomalous one, which is in the low solar corona
(where the solar flare stars) several order of magnitudes higher than the
collisional one. This anomalous resistivity occurs when the drift velocity of
electrons carrying the electric current overcomes some critical velocity
(usually ion-acoustic speed or thermal electron speed --
\citealt{1978A&A....68..145N}, \citealt{1981sfmh.book.....P}). Similar start of
the magnetic reconnection can be expected in all types of stars with hot
corona. Although the initial conditions for the magnetic reconnection in A-type
stars are probably different (due to the missing hot X-ray corona), very
soon after the start of the magnetic reconnection the temperature in the
reconnection site strongly increases. Thus, in the following times the magnetic
reconnection and its spreading into larger volume is in the hot plasma regime.

\begin{figure}[!b]
\includegraphics[width=0.5\textwidth]{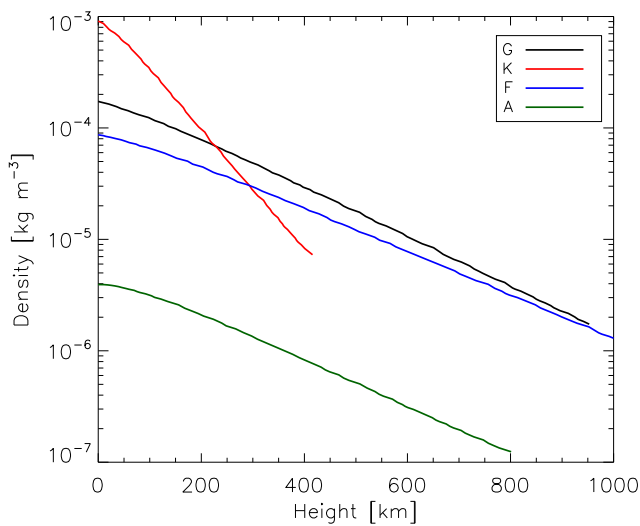}
\caption{Density profiles in the atmospheres of representative stars from MESA models. Optical depth was converted to height above level of unity optical depth using the model of a grey atmosphere.}
\label{fig:atmospheres}
\end{figure}

It is not possible to properly (i.e. including the possible temperature
inversion layers, such as chromosphere or corona, such as the set of VAL,
\citealt{1981ApJS...45..635V}, models in the case of the Sun) model the stellar
atmospheres of the representative stars, hence we stick to the estimates from
the grey atmospheric models computed for optical depths $10^{-4}$ and larger.
We analyzed all plasma parameters of these atmospheres.
The most distinct difference was found in plasma densities. The density on
A-star is more than one order lower than on cooler stars (see
Fig.~\ref{fig:atmospheres}).

Now a question arises what is an effect of lower density at
flare site. Firstly, the radiative losses are much lower, because they are
proportional to the plasma density to the square \citep{1981sfmh.book.....P}.
Thus, owing to this reduction of energy losses higher temperatures in flares
can be reached. Secondly, the threshold for the current-driven instabilities
generating the anomalous resistivity decreases with the density decrease.
Namely, a drift velocity of electrons $v_{\rm D}$ carrying the electric current
is inversely proportional to the electron density $n_{\rm e}$
\begin{equation}
v_{\rm D}=j/(e n_{\rm e}),
\end{equation}
where $j$ is the electric current density and $e$ the electron charge. It means
that the anomalous resistivity can be generated in the current sheet with smaller
derivative of the magnetic field across the current sheet as follows
from
\begin{equation}
j \propto \frac{\partial B}{\partial x},
\end{equation}
where $B$ is the magnetic field component along the current sheet and $x$ is the
coordinate across the current sheet. Namely, in the current sheet the only non-zero derivative of the magnetic
field is that across the current sheet.

A-type stars do not have sub-surface convection zones, however observations
indicate that they must posses photospheric magnetic fields of solar type,
hence with a mixed polarity. This magnetic field probably originates from the
dynamo process occurring in the convective core. Numerical simulations
\citep{2011JPhCS.271a2068F} claim to see a significant dynamo action resulting
even in super-equipartition magnetic fields in the interior of not only A-type
stars, but also on B-type stars \citep{2016arXiv160303659A}. A
super-equipartition (with respect to the convective motion) field may become
unstable, buoyant, and rise through a radiative envelope towards stellar
photosphere. There are no significant bulk motions in this radiative envelope,
hence the field rises up basically in form of the loops and is not dispersed by
convective motions. Since the gas density decreases as does gas pressure, the
field fans out in order to maintain the pressure equilibrium: It becomes
spatially larger, however with smaller intensity. It is analogous to the loops
in the solar atmosphere, fanning out from active regions. Therefore we may
expect the appearance of the magnetic fields with mixed
polarity that have a larger extent.

The observational evidence for large spots on A stars is problematic at least.
\cite{2013MNRAS.431.2240B} estimated the sizes of spots
responsible for the rotational modulation of A-star light
curves and found that these spots are comparable in size with those on the Sun.
This contradicts not only our expectations formulated above, but also findings
with regards to G-type stars \citep{2016arXiv160101132M}, where large spots
(covering 1\% of stellar surface and more) are a necessary condition for
superflares to ignite.

It is possible that the spot areas measured by \cite{2013MNRAS.431.2240B} are
severely underestimated from three reasons: First, the methodology assumes that
the light-curve variations are due to a dark round spot. Then the fractional
decrease in bolometric flux is proportional to the fractional area of the spot
with respect to the visible hemisphere of the star. In reality, these spots are
never completely dark, the relative contrast of sunspots is larger for sunspots
with a stronger magnetic field. We assumed above that the intensity of the
magnetic field in the photosphere of A stars is lower ($\sim 100$~G) than on the Sun ($\sim 1000$~G), hence one may safely
assume that the relative contrast of spots and the surrounding photosphere be
also lower. In effect, this leads to an underestimate of size of the spot.
Second, the method is not sensitive to spots that remain on a visible
hemisphere of the star, e.g. to spots in larger latitudes of star with inclined
axis of rotation. And third, \cite{2011A&A...534A.140C} pointed out a
possibility of a weak dynamo action even in near-surface convective shells of
early-type stars. Such a dynamo would result in brighter spots, similarly to
faculae on the Sun. These bright spots would compensate for some of the darkening caused by dark spots
in the light-curve modulated by rotation and again lead to underestimation of the spot size.

Numerical simulations \citep[e.g.][]{2013A&A...558A..49B,2015A&A...581A..42B}
suggest that a typical scale of convection increases with increasing surface
temperature from around 1~Mm for Sun-like stars to $\sim5$~Mm for a F3 star and
$\sim15$~Mm for an A star \citep{2012JCoPh.231..919F}. The magnetic flux is
dispersed to the same (and larger) scales, which provides an indirect hint for
large-scale magnetic fields to exist in photospheres of A stars.

Thanks to the decreased profile of electron density compared to the F--K stars
the flare may ignite in regions closer to the photosphere of the A star, where
we assume that the intensity of the magnetic field is $\sim100$~G (which is comparable
to the magnetic field intensity at the reconnection site of solar flares). The flare will then engulf a
large volume of the magnetic field resulting in a flare with a large energy
released in total. Since the incidence of flares per star is lower for A-type
stars than for cooler types (there is on average 2.33 flares per star for
A-type, 4.20 for F-type, 16.83 for G-type, and 22.34 for K-type) is would also
seem that it takes a larger period of time for the dynamo to create such a
large magnetic field. Theoretical considerations for the case of our Sun
\citep[e.g.][]{2013PASJ...65...49S} also show that the occurrence rate of large
solar flares is limited by the finite efficiency of the dynamo process.

\section{(Non-)magnetic A-type stars}
There is a tentative evidence for Sun-like flares occurring also on A-type
stars. The Sun-like flares need complex topologies of active regions. There is
no observational evidence for existence of such magnetic fields in atmospheres
of A stars. A well known group of A stars does posses observed magnetic fields.
Ap stars are peculiar stars of type A which show overabundances of some rare
metals. They also have strong magnetic fields typically reaching up to tens of
kG. In most cases a field which is dipole-like and is not aligned with a
rotation axis. It is generally assumed that the the magnetic field of Ap stars
is the fossil in origin, in which the field is a relic of the initial field in
the interstellar medium. The estimated diffusion time for such a magnetic field
in the conditions in the interior of a typical A star is $\sim 10^9$  years
\citep{2004Natur.431..819B}.

None of the flaring A stars in our sample belongs to the group of Ap stars. It
is not surprising. We have mentioned a generally accepted hypothesis that in
order to ignite a Sun-like flare a region with high current density is needed
to exist. The bipolar global magnetic field does not provide conditions for
such high-current-density regions to exist. Therefore one should not expect a
typical Ap star to flare.

On the other hand, \cite{2011JPhCS.271a2068F} performed a fully 3-D
magnetohydrodynamic modelling of a $2M_\odot$ A star with included fossil
magnetic field having a twisted toroidal shape with a strong dipolar component.
The dynamo takes place in the convective core which builds up a strong mean
field and with excursions towards super-equipartition intensities with respect
to the convection. The authors speculated that these fields might possibly
become buoyant (due to the diffusive limitations in the code such a rise was
not observed during various runs), ultimately contributing to form large spots
\citep{2011PhDT.........1F}. Due to the feedback of the magnetic field on
convection, there is almost no latitudinal differential rotation in the core of
such a star. The results of this simulation interestingly fit to a scenario by
\cite{2014IAUS..302..338L}. The authors note that there is a magnetic gap
between A stars with measured large-scale magnetic fields stronger than a few
hundreds Gauss (around 5--10\% of A stars) and A-stars of
Vega-type, where only weak ($\sim$1~G) magnetic fields were measured
spectroscopically. The authors point out that if a magnetic field is stronger
than some 300~G, it is strong enough to avoid macroscopic mixing through
differential rotation, thermal convection, or stellar winds, and the star
becomes a Ap star with chemical peculiarity. Weaker fields do not prevent the
mixing and cannot prevent their winding-up by differential rotation into strong
toroidal fields. Such configuration of the magnetic field is expected to become
unstable (e.g. a Tayler instability, \citealt{1973MNRAS.161..365T}), which
transforms the initially large-scale field configuration into a new
configuration with mixed polarities at the length scale of the instability. On
the contrary, if the initial magnetic field is strong enough, Maxwell stresses
impose uniform rotation and eventually lead to a stable configuration. The
critical field intensity discriminating the two different outputs of coupling
between plasma streaming and magnetic field inside the A star is around 300~G.
This value was experimentally established earlier by
\cite{2007A&A...475.1053A}, who studied spectroscopically a sample of 28 Ap/Bp
stars. They inferred the intensity of the dipole magnetic field and found that
its histogram shows a peak around 1000~G falling off to both smaller and larger
fields. There were no stars with dipole field strength less than 300~G. The
authors claim that their results is neither detection threshold effect nor
selection effect.

We speculate that flaring A stars underwent Tayler instability in the past,
which transformed their initially dipole fossil fields into organised toroidal
ones. These localised fields possibly contain patches of both polarities, as
solar active regions do. Therefore, they do not contribute to a polarimetric
signal, because the disc-integrated polarimetric signals largely cancel out.
The localised magnetic fields in photospheres of such stars would not be measurable using
standard spectroscopic methods.

\section{Implications for coronal heating}
The presence of stellar coronae is traditionally connected with the radiation
in the band of X-rays, due to the expected high temperature of a few million
degrees. X-ray fluxes were studied in several survey projects, namely in ROSAT
all-sky survey. Several authors used these measurements to cross-identify the
X-ray sources with e.g. stellar catalogues. It was found
\citep{1998A&AS..132..155H,2007A&A...475..677S} that the ratio of stars with
detected X-rays and all stars of the given spectral type is not uniform among
spectral types, e.g. there is a prominent underabundance of X-ray-active stars
between spectral types B3 and F0 \citep[see Fig.~2 of][]{2007A&A...475..677S}.
On the high-temperature wall of this `hole' the indications of X-rays
coming from instabilities in the radiatively driven winds on
early-B and hotter stars are visible, on the low-temperature wall the X-rays
increase due to the expected onset of $\alpha\Omega$ dynamo on F stars, related
to the onset of near-surface convection zone. The incidence of X-rays among
A-type stars is less than 20\%, which may be interpreted as non-existence of
hot stellar coronae around A-stars.

The different statistics of occurrence rates of flares on A-type stars compared
to cooler types provides hints to speculate about the corona formation. The
existence of the solar corona has not been successfully explained since almost
75 years \citep{2015RSPTA.37340269D}. One class of models of coronal heating
assumes that flares play a significant role \citep{1988ApJ...330..474P}.
Energetic considerations convincingly show that for the Sun,
the energy released by large solar flares is not sufficient to cover radiative
losses from the 1-million degree corona. There is another evidence that the
heating should be more-or-less stationary, which again precludes large flares
occurring at most once a day to play a significant role. It is therefore assumed
that the role of heating events is played by small-scale flares occurring more
frequently, in general termed nanoflares with energies around $10^{17}$~J. If
the index of the power law in relation (\ref{eq:powerlaw}) is $-2$ or smaller,
then the total energy released by flares is dominated by low-energy events. For
the Sun the power-law index is close to $-2$.

From our results it seems that the power-law index is similar also for F-, and
K-type stars, hence the existence of the hot corona around these stars may be
justified due to the small flares occurring there. However, for the A-type
stars, the power-law is not that steep, around $-1.4$, hence the energy budget
released by flares is dominated by large-energy events. By extrapolating the
fits in the histograms in Fig.~\ref{fig:histograms} towards low-energy part and
integrating the spectrum from nanoflare energies $10^{17}$~J to superflare
energies $10^{30}$~J we obtain estimates for energy flux into the higher
atmosphere by eruptive events. We found that the total integrated flux
(with 1$\sigma$ bounds) is $(6.7-6.8)\times10^{21}$~W for the representative
G-type star, $(3.9-5.6)\times10^{21}$ for the K-type, and
$(1.4-1.6)\times10^{21}$ for the F-type star. Even though
these number are very rough estimates, they are not far from e.g. a quiet-Sun
X-ray luminosity of around $5\times10^{20}$ \citep{1978ARA&A..16..393V}
measured for the Sun. For the representative A-star the energy flux by flare
events using the same methodology is $(6.2-7.3)\times10^{21}$~W. Taking into
account that these stars have a larger surface area, the energy flux density
into higher atmosphere is 4- to 5-times larger in case of the
representative G-type star than in case of the A-type star.

There are two noticeable issues. First, the slope determined for G-type stars is
also significantly lower than critical value of $-2$, distinguishing the dominance of the contribution by small-energy flares and large-energy events. It is evident in the upper panel of Fig.~\ref{fig:histograms} that the shape of the histogram for G stars around the location of the maximum occurrence rate is different from histograms for K and F stars. The width of the maximum of the G-star histogram is much larger than in the other two cases. This fact together with the automatic selection of the low-energy cut-off at the position of the histogram maximum leads to a shallower slope of the fitted power law. Increasing the low-energy cut-off to e.g. $10^{28}$~J leads to a derived slope of ($-1.70\pm0.03$) and the extrapolated total flux by flare events $(7.0-7.8)\times10^{21}$~J. The energy density would then be 5- to 6-times larger than for the A stars. The reason for existence of a pseudo-plateau in the histogram of flares in G stars is unknown and deserves a future investigation.

The second issue is that our estimates are based on observations of superflare stars, whereas our Sun does have a hot corona and it does not seem to be a superflare stars. There are indications that the lack of superflares on our Sun may simply be a consequence of short observational series available for analysis. \cite{2013PASJ...65...49S} showed that the occurence rates for superflares on \kepler G-type stars group in the occurence rate diagram around the same power law as do the occurence rates for solar flares of a large range of energies from nanoflares to regular solar flares (see their Figure 1). Simplistic estimates made by the authors also do not exclude possibility of superflares occuring on our Sun, e.g. $10^{28}$~J flare could on average happen once in 5000~years.

It has to be noted that our speculative claims are based on extrapolation and
a silent assumption that the occurrence rate of flares follows the same power
law over many order of magnitude. With a current instrumentation it is not possible
to verify our ideas directly by observations. However, there are many pieces of
a puzzle described above that seem to fit together to draw a physically sound
picture of the origin of the flares and hot corona formation. These pieces
of evidence include both observations and state-of-the-art numerical simulation,
thereby being largely independent.

The apparently lower energy flux by flaring events may easily explain the lack
for evidence for the hot X-ray emitting coronae on A-type stars and
strengthen the role of nanoflares in heating the corona of the Sun.

\section{Summary}
In agreement with previous studies, we assume that in the A-type flare stars
the magnetic field is amplified by dynamo processes in the
convective core of the stars. Then part of the generated
magnetic field becomes unstable, buoyant and rises as magnetic ropes through  a
radiative envelope towards stellar photosphere.  During this process the
magnetic ropes become larger and their magnetic field decreases. Based on the
flare statistics  we propose that this process generates more large-scale
structures and less small-scale structures of the magnetic field than in the
case of Sun-like stars. Owing to the relative shortage of the small-scale
structures, a number of small flares (nanoflares) is relatively small and the
upper atmospheric layers of the A-like flare star is not sufficiently heated
and  thus {a hot X-ray emitting corona} cannot be formed. On the other hand,
the large-scale structures of the magnetic field in large volumes can be
dissipated in huge flares.

By modelling the stellar atmosphere we show that the gas density in the
atmospheres of A stars is lower, hence the radiative losses are also much
lower. Owing to this reduction of energy losses higher temperatures in flares
can be reached. The lower densities in the photosphere and in the above laying
layers in the A-type flare stars comparing to cooler stars, the threshold for
generation of the anomalous resistivity, which is important in any flares, can
be reached for smaller magnetic field derivatives in flare current sheets.

Contrary to the A-type flare stars, in the Sun and cooler stars the hot
corona is formed. For cooler stars the power-law index of the flare occurrence
looks to be steeper than that for the A-type flare stars, which shows
relatively more nanoflares in these stars. Moreover, it indicates an importance
of nanoflares for heating and formation of the solar corona.

\acknowledgements We are very grateful to Luis Balona (South African
Astronomical Observatory) for making the measurements of stellar flares public
through VizieR interface. The authors were supported by the institute research
project RVO:67985815 to Astronomical Institute of Czech Academy of Sciences.
M.\v{S}. further acknowledges a support from the grant 15-02112S, M.K. from
grant P209/12/0103, both grants were awarded by the Czech Science Foundation.
The original data are based on observations collected by the Kepler mission.
Funding for the Kepler mission is provided by the NASA Science Mission
directorate. These observations were obtained from the Mikulski Archive for
Space Telescopes (MAST). STScI is operated by the Association of Universities
for Research in Astronomy, Inc., under NASA contract NAS5-26555. Support for
MAST for non-HST data is provided by the NASA Office of Space Science via grant
NNX09AF08G and by other grants and contracts." We thank Sacha Brun for useful advices. 
We thank the anonymous referee for useful comments that improved the paper. 


\begin{thebibliography}{}
\expandafter\ifx\csname natexlab\endcsname\relax\def\natexlab#1{#1}\fi

\bibitem[{{Aschwanden} {et~al.}(2000){Aschwanden}, {Tarbell}, {Nightingale},
  {Schrijver}, {Title}, {Kankelborg}, {Martens}, \&
  {Warren}}]{2000ApJ...535.1047A}
{Aschwanden}, M.~J., {Tarbell}, T.~D., {Nightingale}, R.~W., {et~al.} 2000,
  \apj, 535, 1047

\bibitem[{{Aschwanden} {et~al.}(2016){Aschwanden}, {Crosby}, {Dimitropoulou},
  {Georgoulis}, {Hergarten}, {McAteer}, {Milovanov}, {Mineshige}, {Morales},
  {Nishizuka}, {Pruessner}, {Sanchez}, {Sharma}, {Strugarek}, \&
  {Uritsky}}]{2016SSRv..198...47A}
{Aschwanden}, M.~J., {Crosby}, N.~B., {Dimitropoulou}, M., {et~al.} 2016, \ssr,
  198, 47

\bibitem[{{Augustson} {et~al.}(2013){Augustson}, {Brun}, \&
  {Toomre}}]{2013ApJ...777..153A}
{Augustson}, K.~C., {Brun}, A.~S., \& {Toomre}, J. 2013, \apj, 777, 153

\bibitem[{{Augustson} {et~al.}(2016){Augustson}, {Brun}, \&
  {Toomre}}]{2016arXiv160303659A}
---. 2016, ArXiv e-prints, arXiv:1603.03659

\bibitem[{{Auri{\`e}re} {et~al.}(2007){Auri{\`e}re}, {Wade}, {Silvester},
  {Ligni{\`e}res}, {Bagnulo}, {Bale}, {Dintrans}, {Donati}, {Folsom},
  {Gruberbauer}, {Hui Bon Hoa}, {Jeffers}, {Johnson}, {Landstreet},
  {L{\`e}bre}, {Lueftinger}, {Marsden}, {Mouillet}, {Naseri}, {Paletou},
  {Petit}, {Power}, {Rincon}, {Strasser}, \& {Toqu{\'e}}}]{2007A&A...475.1053A}
{Auri{\`e}re}, M., {Wade}, G.~A., {Silvester}, J., {et~al.} 2007, \aap, 475,
  1053

\bibitem[{{Baliunas} {et~al.}(1995){Baliunas}, {Donahue}, {Soon}, {Horne},
  {Frazer}, {Woodard-Eklund}, {Bradford}, {Rao}, {Wilson}, {Zhang}, {Bennett},
  {Briggs}, {Carroll}, {Duncan}, {Figueroa}, {Lanning}, {Misch}, {Mueller},
  {Noyes}, {Poppe}, {Porter}, {Robinson}, {Russell}, {Shelton}, {Soyumer},
  {Vaughan}, \& {Whitney}}]{1995ApJ...438..269B}
{Baliunas}, S.~L., {Donahue}, R.~A., {Soon}, W.~H., {et~al.} 1995, \apj, 438,
  269

\bibitem[{{Balona}(2012)}]{2012MNRAS.423.3420B}
{Balona}, L.~A. 2012, \mnras, 423, 3420

\bibitem[{{Balona}(2013)}]{2013MNRAS.431.2240B}
---. 2013, \mnras, 431, 2240

\bibitem[{{Balona}(2015)}]{2015MNRAS.447.2714B}
---. 2015, \mnras, 447, 2714

\bibitem[{{Bauke}(2007)}]{2007EPJB...58..167B}
{Bauke}, H. 2007, European Physical Journal B, 58, 167

\bibitem[{{Beeck} {et~al.}(2013){Beeck}, {Cameron}, {Reiners}, \&
  {Sch{\"u}ssler}}]{2013A&A...558A..49B}
{Beeck}, B., {Cameron}, R.~H., {Reiners}, A., \& {Sch{\"u}ssler}, M. 2013,
  \aap, 558, A49

\bibitem[{{Beeck} {et~al.}(2015){Beeck}, {Sch{\"u}ssler}, {Cameron}, \&
  {Reiners}}]{2015A&A...581A..42B}
{Beeck}, B., {Sch{\"u}ssler}, M., {Cameron}, R.~H., \& {Reiners}, A. 2015,
  \aap, 581, A42

\bibitem[{{Braithwaite} \& {Spruit}(2004)}]{2004Natur.431..819B}
{Braithwaite}, J., \& {Spruit}, H.~C. 2004, \nat, 431, 819

\bibitem[{{Candelaresi} {et~al.}(2014){Candelaresi}, {Hillier}, {Maehara},
  {Brandenburg}, \& {Shibata}}]{2014ApJ...792...67C}
{Candelaresi}, S., {Hillier}, A., {Maehara}, H., {Brandenburg}, A., \&
  {Shibata}, K. 2014, \apj, 792, 67

\bibitem[{{Cantiello} \& {Braithwaite}(2011)}]{2011A&A...534A.140C}
{Cantiello}, M., \& {Braithwaite}, J. 2011, \aap, 534, A140

\bibitem[{{Collier Cameron}(1996)}]{1996IAUS..176..449C}
{Collier Cameron}, A. 1996, in IAU Symposium, Vol. 176, Stellar Surface
  Structure, ed. K.~G. {Strassmeier} \& J.~L. {Linsky}, 449

\bibitem[{{Crosby} {et~al.}(1993){Crosby}, {Aschwanden}, \&
  {Dennis}}]{1993SoPh..143..275C}
{Crosby}, N.~B., {Aschwanden}, M.~J., \& {Dennis}, B.~R. 1993, \solphys, 143,
  275

\bibitem[{{De Moortel} \& {Browning}(2015)}]{2015RSPTA.37340269D}
{De Moortel}, I., \& {Browning}, P. 2015, Philosophical Transactions of the
  Royal Society of London Series A, 373, 20140269

\bibitem[{{Featherstone}(2011)}]{2011PhDT.........1F}
{Featherstone}, N.~A. 2011, PhD thesis, University of Colorado at Boulder

\bibitem[{{Featherstone} {et~al.}(2011){Featherstone}, {Browning}, {Brun}, \&
  {Toomre}}]{2011JPhCS.271a2068F}
{Featherstone}, N.~A., {Browning}, M.~K., {Brun}, A.~S., \& {Toomre}, J. 2011,
  Journal of Physics Conference Series, 271, 012068

\bibitem[{{Freytag} {et~al.}(2012){Freytag}, {Steffen}, {Ludwig},
  {Wedemeyer-B{\"o}hm}, {Schaffenberger}, \& {Steiner}}]{2012JCoPh.231..919F}
{Freytag}, B., {Steffen}, M., {Ludwig}, H.-G., {et~al.} 2012, Journal of
  Computational Physics, 231, 919

\bibitem[{{Huensch} {et~al.}(1998){Huensch}, {Schmitt}, \&
  {Voges}}]{1998A&AS..132..155H}
{Huensch}, M., {Schmitt}, J.~H.~M.~M., \& {Voges}, W. 1998, \aaps, 132, 155

\bibitem[{{Kallinger} \& {Matthews}(2010)}]{2010ApJ...711L..35K}
{Kallinger}, T., \& {Matthews}, J.~M. 2010, \apjl, 711, L35

\bibitem[{{Ligni{\`e}res} {et~al.}(2014){Ligni{\`e}res}, {Petit},
  {Auri{\`e}re}, {Wade}, \& {B{\"o}hm}}]{2014IAUS..302..338L}
{Ligni{\`e}res}, F., {Petit}, P., {Auri{\`e}re}, M., {Wade}, G.~A., \&
  {B{\"o}hm}, T. 2014, in IAU Symposium, Vol. 302, Magnetic Fields throughout
  Stellar Evolution, ed. P.~{Petit}, M.~{Jardine}, \& H.~C. {Spruit}, 338--347

\bibitem[{{Ligni{\`e}res} {et~al.}(2009){Ligni{\`e}res}, {Petit}, {B{\"o}hm},
  \& {Auri{\`e}re}}]{2009A&A...500L..41L}
{Ligni{\`e}res}, F., {Petit}, P., {B{\"o}hm}, T., \& {Auri{\`e}re}, M. 2009,
  \aap, 500, L41

\bibitem[{{Maehara} {et~al.}(2016){Maehara}, {Shibayama}, {Notsu}, {Notsu},
  {Honda}, {Nogami}, \& {Shibata}}]{2016arXiv160101132M}
{Maehara}, H., {Shibayama}, T., {Notsu}, Y., {et~al.} 2016, ArXiv e-prints,
  arXiv:1601.01132

\bibitem[{{Maehara} {et~al.}(2012){Maehara}, {Shibayama}, {Notsu}, {Notsu},
  {Nagao}, {Kusaba}, {Honda}, {Nogami}, \& {Shibata}}]{2012Natur.485..478M}
{Maehara}, H., {Shibayama}, T., {Notsu}, S., {et~al.} 2012, \nat, 485, 478

\bibitem[{{Narain} \& {Ulmschneider}(1996)}]{1996SSRv...75..453N}
{Narain}, U., \& {Ulmschneider}, P. 1996, \ssr, 75, 453

\bibitem[{{Norman} \& {Smith}(1978)}]{1978A&A....68..145N}
{Norman}, C.~A., \& {Smith}, R.~A. 1978, \aap, 68, 145

\bibitem[{{Parker}(1988)}]{1988ApJ...330..474P}
{Parker}, E.~N. 1988, \apj, 330, 474

\bibitem[{{Paxton} {et~al.}(2011){Paxton}, {Bildsten}, {Dotter}, {Herwig},
  {Lesaffre}, \& {Timmes}}]{2011ApJS..192....3P}
{Paxton}, B., {Bildsten}, L., {Dotter}, A., {et~al.} 2011, \apjs, 192, 3

\bibitem[{{Petit} {et~al.}(2010){Petit}, {Ligni{\`e}res}, {Wade},
  {Auri{\`e}re}, {B{\"o}hm}, {Bagnulo}, {Dintrans}, {Fumel}, {Grunhut},
  {Lanoux}, {Morgenthaler}, \& {Van Grootel}}]{2010A&A...523A..41P}
{Petit}, P., {Ligni{\`e}res}, F., {Wade}, G.~A., {et~al.} 2010, \aap, 523, A41

\bibitem[{{Pettersen}(1989)}]{1989SoPh..121..299P}
{Pettersen}, B.~R. 1989, \solphys, 121, 299

\bibitem[{{Priest}(1981)}]{1981sfmh.book.....P}
{Priest}, E.~R. 1981, {Solar flare magnetohydrodynamics}

\bibitem[{{Rauer} {et~al.}(2014){Rauer}, {Catala}, {Aerts}, {Appourchaux},
  {Benz}, {Brandeker}, {Christensen-Dalsgaard}, {Deleuil}, {Gizon}, {Goupil},
  {G{\"u}del}, {Janot-Pacheco}, {Mas-Hesse}, {Pagano}, {Piotto}, {Pollacco},
  {Santos}, {Smith}, {Su{\'a}rez}, {Szab{\'o}}, {Udry}, {Adibekyan}, {Alibert},
  {Almenara}, {Amaro-Seoane}, {Eiff}, {Asplund}, {Antonello}, {Barnes},
  {Baudin}, {Belkacem}, {Bergemann}, {Bihain}, {Birch}, {Bonfils}, {Boisse},
  {Bonomo}, {Borsa}, {Brand{\~a}o}, {Brocato}, {Brun}, {Burleigh}, {Burston},
  {Cabrera}, {Cassisi}, {Chaplin}, {Charpinet}, {Chiappini}, {Church},
  {Csizmadia}, {Cunha}, {Damasso}, {Davies}, {Deeg}, {D{\'{\i}}az}, {Dreizler},
  {Dreyer}, {Eggenberger}, {Ehrenreich}, {Eigm{\"u}ller}, {Erikson}, {Farmer},
  {Feltzing}, {de Oliveira Fialho}, {Figueira}, {Forveille}, {Fridlund},
  {Garc{\'{\i}}a}, {Giommi}, {Giuffrida}, {Godolt}, {Gomes da Silva},
  {Granzer}, {Grenfell}, {Grotsch-Noels}, {G{\"u}nther}, {Haswell}, {Hatzes},
  {H{\'e}brard}, {Hekker}, {Helled}, {Heng}, {Jenkins}, {Johansen},
  {Khodachenko}, {Kislyakova}, {Kley}, {Kolb}, {Krivova}, {Kupka}, {Lammer},
  {Lanza}, {Lebreton}, {Magrin}, {Marcos-Arenal}, {Marrese}, {Marques},
  {Martins}, {Mathis}, {Mathur}, {Messina}, {Miglio}, {Montalban}, {Montalto},
  {Monteiro}, {Moradi}, {Moravveji}, {Mordasini}, {Morel}, {Mortier},
  {Nascimbeni}, {Nelson}, {Nielsen}, {Noack}, {Norton}, {Ofir}, {Oshagh},
  {Ouazzani}, {P{\'a}pics}, {Parro}, {Petit}, {Plez}, {Poretti}, {Quirrenbach},
  {Ragazzoni}, {Raimondo}, {Rainer}, {Reese}, {Redmer}, {Reffert},
  {Rojas-Ayala}, {Roxburgh}, {Salmon}, {Santerne}, {Schneider}, {Schou},
  {Schuh}, {Schunker}, {Silva-Valio}, {Silvotti}, {Skillen}, {Snellen}, {Sohl},
  {Sousa}, {Sozzetti}, {Stello}, {Strassmeier}, {{\v S}vanda}, {Szab{\'o}},
  {Tkachenko}, {Valencia}, {Van Grootel}, {Vauclair}, {Ventura}, {Wagner},
  {Walton}, {Weingrill}, {Werner}, {Wheatley}, \&
  {Zwintz}}]{2014ExA....38..249R}
{Rauer}, H., {Catala}, C., {Aerts}, C., {et~al.} 2014, Experimental Astronomy,
  38, 249

\bibitem[{{Ricker} {et~al.}(2015){Ricker}, {Winn}, {Vanderspek}, {Latham},
  {Bakos}, {Bean}, {Berta-Thompson}, {Brown}, {Buchhave}, {Butler}, {Butler},
  {Chaplin}, {Charbonneau}, {Christensen-Dalsgaard}, {Clampin}, {Deming},
  {Doty}, {De Lee}, {Dressing}, {Dunham}, {Endl}, {Fressin}, {Ge}, {Henning},
  {Holman}, {Howard}, {Ida}, {Jenkins}, {Jernigan}, {Johnson}, {Kaltenegger},
  {Kawai}, {Kjeldsen}, {Laughlin}, {Levine}, {Lin}, {Lissauer}, {MacQueen},
  {Marcy}, {McCullough}, {Morton}, {Narita}, {Paegert}, {Palle}, {Pepe},
  {Pepper}, {Quirrenbach}, {Rinehart}, {Sasselov}, {Sato}, {Seager},
  {Sozzetti}, {Stassun}, {Sullivan}, {Szentgyorgyi}, {Torres}, {Udry}, \&
  {Villasenor}}]{2015JATIS...1a4003R}
{Ricker}, G.~R., {Winn}, J.~N., {Vanderspek}, R., {et~al.} 2015, Journal of
  Astronomical Telescopes, Instruments, and Systems, 1, 014003

\bibitem[{{Schaefer} {et~al.}(2000){Schaefer}, {King}, \&
  {Deliyannis}}]{2000ApJ...529.1026S}
{Schaefer}, B.~E., {King}, J.~R., \& {Deliyannis}, C.~P. 2000, \apj, 529, 1026

\bibitem[{{Schr{\"o}der} \& {Schmitt}(2007)}]{2007A&A...475..677S}
{Schr{\"o}der}, C., \& {Schmitt}, J.~H.~M.~M. 2007, \aap, 475, 677

\bibitem[{{Shibata} \& {Magara}(2011)}]{2011LRSP....8....6S}
{Shibata}, K., \& {Magara}, T. 2011, Living Reviews in Solar Physics, 8,
  doi:10.12942/lrsp-2011-6

\bibitem[{{Shibata} {et~al.}(2013){Shibata}, {Isobe}, {Hillier}, {Choudhuri},
  {Maehara}, {Ishii}, {Shibayama}, {Notsu}, {Notsu}, {Nagao}, {Honda}, \&
  {Nogami}}]{2013PASJ...65...49S}
{Shibata}, K., {Isobe}, H., {Hillier}, A., {et~al.} 2013, \pasj, 65,
  arXiv:1212.1361

\bibitem[{{Shibayama} {et~al.}(2013){Shibayama}, {Maehara}, {Notsu}, {Notsu},
  {Nagao}, {Honda}, {Ishii}, {Nogami}, \& {Shibata}}]{2013ApJS..209....5S}
{Shibayama}, T., {Maehara}, H., {Notsu}, S., {et~al.} 2013, \apjs, 209, 5

\bibitem[{{Shimizu}(1995)}]{1995PASJ...47..251S}
{Shimizu}, T. 1995, \pasj, 47, 251

\bibitem[{{Stello} {et~al.}(2016){Stello}, {Cantiello}, {Fuller}, {Huber},
  {Garc{\'{\i}}a}, {Bedding}, {Bildsten}, \& {Silva
  Aguirre}}]{2016Natur.529..364S}
{Stello}, D., {Cantiello}, M., {Fuller}, J., {et~al.} 2016, \nat, 529, 364

\bibitem[{{Strassmeier}(2009)}]{2009A&ARv..17..251S}
{Strassmeier}, K.~G. 2009, \aapr, 17, 251

\bibitem[{{Tayler}(1973)}]{1973MNRAS.161..365T}
{Tayler}, R.~J. 1973, \mnras, 161, 365

\bibitem[{{Vaiana} \& {Rosner}(1978)}]{1978ARA&A..16..393V}
{Vaiana}, G.~S., \& {Rosner}, R. 1978, \araa, 16, 393

\bibitem[{{Vernazza} {et~al.}(1981){Vernazza}, {Avrett}, \&
  {Loeser}}]{1981ApJS...45..635V}
{Vernazza}, J.~E., {Avrett}, E.~H., \& {Loeser}, R. 1981, \apjs, 45, 635

\bibitem[{{Wilson}(1978)}]{1978ApJ...226..379W}
{Wilson}, O.~C. 1978, \apj, 226, 379

\end{thebibliography}

\end{document}